\documentstyle[preprint,eqsecnum,aps]{revtex}
\begin{document}
\draft
\parindent=20pt

\title {$\Lambda$-$\Sigma$ conversion in $^4_{\Lambda}$He and
$^4_{\Lambda}$H based on four-body calculation}

\author{E. Hiyama}

\address{Institute of Particle and Nuclear Studies,
High Energy Accelerator Research Organization (KEK),
Tsukuba, 305-0801, Japan} 

\author{M. Kamimura}

\address{Department of Physics, Kyushu University, 
Fukuoka 812-8581,Japan}
 
\author{T. Motoba}

\address{Laboratory of Physics, Osaka Electro-Comm.
University, Neyagawa 572-8530, Japan}

\author{T. Yamada}

\address{Laboratory of Physics, Kanto Gakuin University, 
Yokohama 236-8501, Japan}

\author{Y. Yamamoto}

\address{Physics Section, Tsuru University, Tsuru, Yamanashi 402-8555, 
Japan}
%
\maketitle
\begin{abstract}
Precise four-body calculations for $^4_{\Lambda}$He and
$^4_{\Lambda}$H have been performed in the framework of
the variational method with
Jacobian-coordinate Gaussian-basis functions.
All the rearrangement channels of
both  $NNN \Lambda$ and $NNN \Sigma$  are
explicitly taken into account for the first time 
with the use of realistic $NN$ and $YN$
interactions.
The role of $\Lambda$-$\Sigma$ conversion and 
the amount of the virtual $\Sigma$-component in 
$^4_{\Lambda}$He and $^4_{\Lambda}$H are discussed.
\end{abstract}

\pacs{21.80.+a,21.10.Dr,21.10.Gv,21.45.+v}%

\vfill\eject

\section{Introduction}
\label{sec:int}
It is one of the most fundamental problems in hypernuclear physics
to extract novel information of $YN$ interactions through precise
calculations for a few-body systems such as
$^3_{\Lambda}$H, $^4_{\Lambda}$He and $^4_{\Lambda}$H.
Although a number of studies for $^4_{\Lambda}$He and
$^4_{\Lambda}$H have been performed with
the use of various models so far,
the relations to the underlying $YN$ interactions
are dependent on the adopted models \cite{Gibson72,Gibson88,Aka00}.
In order to explore 
the features of the $YN$ interactions, it is highly required  to
perform
four-body calculation without any restriction in the
configuration space.
An especially interesting issue in this relation is
to make clear the role of $\Lambda$-$\Sigma$ conversion
and to get a reliable estimate of
the amount of the virtual $\Sigma$-component in $\Lambda$
hypernuclei.

However, the $\Lambda N$-$\Sigma N$ coupling terms of
the $YN$ one-boson-exchange (OBE) models proposed so far
have a lot of ambiguity due to
scarce information from the $YN$ scattering experiments.
Recently, a series of the realistic $YN$ interaction
models called NSC97a$\sim$f \cite{Rij99}
have been proposed in which
the spin-spin strengths are varied within the
acceptable range in view of the limited $YN$ scattering
data. Correspondingly, by including the
$\Lambda N$-$\Sigma N$ coupling explicitly,
Miyagawa {\it et al}. \cite{Miyagawa} performed the Faddeev
calculations for
$^3_{\Lambda}$H extensively to test these OBE potential models
in this lightest system. They found that the
$\Lambda N$-$\Sigma N$  coupling is crucial to get
the bound state of $^3_{\Lambda}$H and 
that only the version-f potential  is
acceptable among NSC97 models.
For the further examination on the 
$\Lambda N$-$\Sigma N$ coupling,
$^4_{\Lambda}$He and $^4_{\Lambda}$H are much more useful 
 because both of the spin-doublet states
have been observed. Regarding the representative 
calculations done so far, Gibson {\it et al}. 
employed the coupled two-body model \cite{Gibson72} 
of $^3$He($^3$H)+$\Lambda/ \Sigma$ which was adopted 
originally by Dalitz and Downs \cite{Dalitz58}, and then 
carried out the four-body coupled-channel calculation with the 
separable potentials of central nature \cite{Gibson88}. 
J.  Carlson tried to perform four-body calclation of
these hypernuclei with NSC89 model by using Variational Monte Carlo
method \cite{Carlson} and calculated  the binding energies
with  statistical errors of 100 keV.
Akaishi {\it et al}. \cite{Aka00} recently analyzed the role of the 
$\Lambda N$-$\Sigma N$ coupling for the $0^+$-$1^+$ splitting also 
in the 
framework of the coupled two-body model of 
$^3$He+$\Lambda/\Sigma$.
Now our concern is to perform precise four-body calculations 
of $NNN \Lambda$ and $NNN \Sigma$ irrespectively of any
model assumptions.

It should be noted that, when one goes from $^3_{\Lambda}$H to the
$A=4$ strange systems by allowing
the $\Lambda$-$\Sigma$ conversion, the
computational difficulty increases tremendously.
Recently we have successfully performed the
extensive four-body calculations without any restriction on the
configuration space:
Both the $NNN \Lambda$ and $NNN \Sigma$ channels 
have been incorporated explicitly
and all the rearrangement channels of these baryons
are taken into account.
The variational method with the use
of Jacobian-coordinate Gaussian-basis functions
\cite{Kami88,Kame89} is adopted here; it 
has been proved to provide us with precise computational
results for few-body systems 
\cite{Kami88,Kame89,Hiyama95,Hiyama96,Hiyama00}.

The main purpose of this work is, first, to
solve four-body problem of $^4_{\Lambda}$He and
$^4_{\Lambda}$H by taking into account the $NNN \Lambda (\Sigma)$
channels explicitly with the use of realistic $NN$ and
$YN$ interactions and, secondly, to clarify the role of 
the $\Lambda N$-$\Sigma N$ coupling in the $A=4$ hypernuclei
quantitatively. 

In Sec. II, we describe our method to solve the four-body
problem and $YN$ and $NN$ interactions adopted here.
In Sec.III, we calculate the $\Lambda$ binding energies,
the $\Sigma$-mixing probabilities and the $\Lambda N$ and
$\Sigma N$ correlation functions in $^4_{\Lambda}$He and
$^4_{\Lambda}$H.
The role of the $\Lambda N$-$\Sigma N$ coupling is
investigated by dividing its whole contribution into
the $\Lambda N$ two-body and $\Lambda NN$ three-body parts.
A summary is given in Sec. IV.

\section{Method and interactions}
\label{sec:rge}

As the first step before going to the use of sophisticated
OBE models, we employ here the $\Lambda N$-$\Sigma N$ 
coupled $YN$ potential 
with central, spin-orbit and tensor terms \cite{Shinmura98}
which simulates the scattering phase shifts given by NSC97f.
The potentials parameters for central, spin-orbit and tensor terms are
listed in Table I.

The main reason of using this simulated version of NSC97f is
to  focus our attention clearly on 
 the physical ingredients  as well as for
computational tractability.
In this relation, our most important criterion for
selecting the $YN$ interaction is that
the observed binding energy of
$^3_{\Lambda}$H is reproduced reasonably:
The interaction used here leads to 
the $\Lambda$-binding
energy, $B_{\Lambda}(^3_{\Lambda}{\rm H}), 0.19$ MeV
which agrees  well with the observed data
$(B_{\Lambda}(^3_{\Lambda}{\rm H})=0.13 \pm 0.05$ MeV).
As for the $NN$ interaction, we employ the AV8 potential
\cite{av8}.
We shall, however, examine how the choice of the $NN$ potential
affect the $\Lambda$ binding energy; we additionally take
the Bonn A potential \cite{Mach} 
and the Minnesota potential (central force only) \cite{Minne} 
and investigate the role of the tensor term of $NN$ potential.

The total four-body 
wavefunction is described as a sum of the amplitudes of
all the rearrangement channels $(c=1-4)$ of Fig. 1 in the $LS$
coupling scheme:  
$$
     \Psi_{JM}(^4_{\Lambda}{\rm He}, ^{4}_{\Lambda}{\rm H})
 =   \sum_{Y=\Lambda,\Sigma} \sum_{c=1}^{4}
       \sum_{\alpha I}
       \sum_{ss'S tt'} 
       C^{(c)}_{\alpha I ss'S tt'}  \\  
$$
\vskip -0.5 true cm
$$ 
\hskip 0.0 true cm  
       \times  {\cal A}  \left\{ 
        \Big[\Phi^{(Y,c)}_{\alpha I}({\bf r}_c,{\bf R}_c,
         \mbox{\boldmath $\rho$}_c)    \right.
          \big[ [\chi_{s'}(12)
        \chi_{\frac{1}{2}}(3)]_s 
        \chi_{\frac{1}{2}}(Y) 
          \big]_{S}
       \Big]_{JM} \\
$$
\vskip -0.7 true cm
$$ 
\hskip 1.0 true cm  
    \left.   \times   
          \big[ [\eta_{t'}(12)
        \eta_{\frac{1}{2}}(3)]_{t} 
        \eta_{t_Y}(Y) 
          \big]_{T={\frac{1}{2}}} \right\} , \qquad \qquad \qquad(1)
$$
\noindent
where the spatial part is expressed,
with a set of quantum numbers
 $\alpha=\{nl, NL, K, \nu \lambda \}$,
 by
$$
        \Phi_{\alpha IM}({\bf r},{\bf R},
         \mbox{\boldmath $\rho$})   
         =
       \Big[ [\phi_{nl}({\bf r}) 
         \psi_{NL}({\bf R})]_K
        \xi_{\nu\lambda} 
         (\mbox{\boldmath $\rho$}) \Big]_{IM}. \\ 
$$
Here, $\cal A$ is the three-nucleon
antisymmetrization operator and 
$\chi$'s and $\eta$'s are the spin and isospin functions,
respectively, 
with the isospin $t_Y=0 \:(1)$ for $Y=\Lambda \:(\Sigma)$. 
 Functional form of 
$\phi_{nlm}({\bf r})$ is taken as
$
      \phi_{nlm}({\bf r})
      =
      r^l \, e^{-(r/r_n)^2} 
       Y_{lm}({\widehat {\bf r}}) 
$
where the Gaussian range parameters are chosen to lie in a
geometrical progression ($ r_n=
      r_1 a^{n-1}; n=1 \sim n_{\rm max}$),
and similarly for $\psi_{NL}({\bf R})$ and
$\xi_{\nu\lambda}(\mbox{\boldmath $\rho$})$.
These basis functions were verified to be suited for
describing both the short-range correlations 
and the long-range tail behaviour of few-body systems
\cite{Kami88,Kame89,Hiyama95,Hiyama96}.
  Eigenenergies of the Hamiltonian and 
coefficients $C$'s are determined by 
the Rayleigh-Ritz variational method. 
The angular momentum space of $l, L, \lambda \leq 2$ is found to
be enough to get sufficient convergence of the
calculated results mentioned below.

\section{Results and discussion}
\label{sec:thr}
   
All the calculations have been performed both for
$^4_{\Lambda}$He and $^4_{\Lambda}$H.
Calculated $B_{\Lambda}$ 
 of the $0^+$ ground state and the
$1^+$ excited state of 
$^4_{\Lambda}$He and $^4_{\Lambda}$H are illustrated in
Fig.2 together with the observed values
(also see Table I).
In the case of taking only the $NNN \Lambda$ channel, both of the two 
states are unbound.
Here, the $NNN \Sigma$ sector is divided into the
$(NNN)_\frac{1}{2} \Sigma$ and $(NNN)_\frac{3}{2} \Sigma$ channels
in which the three nucleons are coupled to isospin $t=1/2$ and $3/2$,
respectively.
When the $(NNN)_\frac{1}{2} \Sigma$ channel 
is included,
the $0^+$ state becomes bound, but the $1^+$ state is still
unbound. Then we found that the $1^+$ state
becomes bound only when the $(NNN)_\frac{3}{2} \Sigma$
channel is switched on. It is  noted that
the binding energy
of the $0^+$ state increases only slightly with this $t=3/2$ channel.
The $\Sigma$-channel components turn out to 
play an essential role in the binding mechanism of the $A=4$
hypernuclei,
the $(NNN)_\frac{3}{2} \Sigma$ 
channel being specially important in the $1^+$ state.
The calculated binding energy of the
$0^+$ state almost reproduces the observed binding energy,
while
the $1^+$ state is less bound  by 0.6 (0.4) MeV for 
$^4_{\Lambda}$H ($^4_{\Lambda}$H), 
and hence the $0^+$-$1^+$ splitting is larger than the observed
splitting.
The calculated value of
$B_{\Lambda}(^4_{\Lambda}{\rm He}(0^+))-B_{\Lambda}(^4_{\Lambda}{\rm
H}(0^+))
=-0.05$ MeV is different from the experimental one, $+0.35$ MeV,
although the Coulomb potentials between charged particles 
($p, \Sigma^{\pm})$ are included.
This difference should be attributed
to the charge-symmetry-breaking component
which is not included in  our adopted $YN$ interaction.

Here, we examine how the calculated $\Lambda$ binding energies
depend on the properties of adopted $NN$ potentials.
Band\={o} {\it et al.} once pointed out that
the rearrangement effect originated from the $NN$ tensor force
plays an important role for the mass dependence of $\Lambda$
binding energies of $A=3 \sim 5$ hypernuclei.
Now let us focus our concern on the $NN$ tensor forces:
The three $NN$ potentials compared here are
the AV8 potential with a stronger tensor force,
the Bonn A potential with a weaker tensor force and
the Minnesota potential with a central force only.
The binding energy of the $_\Lambda^3$H is reproduced
reasonably when these $NN$ potentials are used together
with our $YN$ potential.
In Fig. 3 the calculated $\Lambda$ binding energies
of the $0^+$ and $1^+$ states in $^4_{\Lambda}$He
are shown in the cases of using these $NN$ potentials.
It should be noted here that the larger ratio of
the $NN$ tensor component turns out to bring about
the less binding of $\Lambda$.
In the cases of using AV8 and Bonn A, in which the
tensor compornents are included more or less realistically,
the observed $\Lambda$ binding energy is reproduced well.
However. the use of the effective model with no tensor
force such as the Minnesota potential gives rise to
the overbinding of $\Lambda$ by 0.4 MeV.
Hereafter, we report only the results for the AV8 potential,
because there is no meaningful differece between
the results for the AV8 and Bonn A potentials.

As listed in Table III,
the calculated probabilities of the $NNN \Sigma$-channel admixture
are  2.08\%  and 1.03\% 
for the $0^+$ and $1^+$ states
in $^4_{\Lambda}$He, respectively.
In the $0^+$ state, the probability of the $(NNN)_\frac{1}{2} \Sigma$
channel is much larger than that of the
$(NNN)_\frac{3}{2} \Sigma$ channel, while in the $1^+$ state they
are nearly the same. 
We therefore confirm that the $(NNN)_\frac{3}{2} \Sigma$
channel is especially important in the $1^+$ state.
The $S$-, $P$- and $D$-state probabilities of the
channels are also listed in Table III. 
It is remarkable that, in the $NNN \Sigma$ channel,
the $D$-state component is dominant both in the
$0^+$ and $1^+$ states, 
since the $\Lambda N$-$\Sigma N$ 
coupling part 
of the present interaction is dominated by
the tensor component.
These properties are almost similar  in the case of 
$^4_\Lambda$H.

  It is of interest to see  spatial locations of 
$N, \Lambda$ and $\Sigma$ particles in the $A=4$ hypernuclei
 and the correrations between each pairs.
For the $0^+$ state of $^4_{\Lambda}$He,
we illustrated 
the correlation functions (two-body
densities) of the  $NN$,
$\Lambda N$  and $\Sigma N$ pairs (Fig. 4)
and
the (one-body) densities
of single nucleon and $\Lambda$ and $\Sigma$ hyperons (Fig. 5).
Also, we calculated 
the r.m.s. distances between the pair particles 
($\bar{r}_{NN}, \bar{r}_{\Lambda N},\bar{r}_{\Sigma N}) $
as well as  
the r.m.s. radii $(\bar{r}_N, \bar{r}_\Lambda,
\bar{r}_\Sigma)$ of $N, \Lambda$ and $\Sigma$
measured from the c.m. of  $3N$ (Table II).
In Fig. 4, the $NN$ correlation function  in $^4_{\Lambda}$He
exhibits almost the same shape as that in the $^3$He nucleus,
indicating that the dynamical change of the nuclear size  due to the 
$\Lambda$ participation is small. 
The $\Lambda N$ correlation function is of larger range and  flatter
than the $NN$ one, because the 
strength of the $\Lambda N$ interaction is significantly 
smaller than the $NN$ case.
The $\Sigma N$ correlation function is 
much shorter-ranged than the $\Lambda N$ one due to 
the large virtual excitation energy (80 MeV) of 
$\Lambda \rightarrow$ $\Sigma$.
These features are verified by the r.m.s.
distances ($\bar{r}_{NN}, \bar{r}_{\Lambda N},\bar{r}_{\Sigma N}) $
listed in Table I.

Furthermore, we notice that 
the $\Lambda$ particle is  
located much outside the core nucleons 
(see $\bar{r}_{\Lambda} > \bar{r}_N$ and Fig. 5)
and therefore the dynamical change of the core nucleus 
due to the $\Lambda$-particle partition is small:
The nucleon r.m.s. radius is $\bar{r}_N=1.65$ fm  
which is shrinked by 8~\%
from the corresponding one,
$\bar{r}_N=$ 1.79 fm in $^3$He. 
On the other hand, the $\Sigma$ hyperon 
comes close to the nucleons 
(see $\bar{r}_{\Sigma N} < \bar{r}_{\Lambda N}$ and Fig. 4)
and therefore generates a large dynamical 
contraction of the core nucleus in the
$NNN\Sigma$-channel space;
$\bar{r}_N=$  1.49 fm is obtained 
with the $\Sigma$-channel amplitude only
and
the reduction rate amounts to  $17~\%$.
Furthermore, it is interesting to see  in Fig. 4 
the following feature of
the $\Sigma$-admixture effect: In spite of
the totally small probability of
the $\Sigma$-mixing (2 \%),
the $\Sigma N$ components at short distances are not so small
in comparison with the $\Lambda N$ ones.
This enhanced short-ranged component of the $\Sigma$-mixing is 
expected to be reflected in the non-mesic decay of
$\Sigma N \rightarrow NN$.

In order to show the physical effect of 
$\Sigma$-mixing in more detail,
let us separate the whole contribution of 
the $\Lambda N$-$\Sigma N$ coupling interaction into the 
following two processes illustrated in Fig. 6:
The first one is the process (i) which can be renormalized into the
effective $\Lambda N$ two-body force and the
second one is the process (ii) which can be represented by the 
effective $\Lambda NN$ three-body force acting in the
$NNN \Lambda$ space.
We solve the Schr\"{o}dinger equation
by excluding the three-body process (ii) so as to
evaluate the contribution of the
process (i) only.  As shown in Fig. 6,
the process (i) is
large enough to make both the $0^+$ and $1^+$ states bound.
We found that the three-body process(ii) is substantial:
This process results in attraction by 0.62 MeV in the
$0^+$ state, while repulsion by 0.09 MeV
in the $1^+$ state of $^4_{\Lambda}$He and similarly for $^4_{\Lambda}$H. 

In order to investigate the respective role of the central and the 
tensor terms of the $\Lambda N$-$\Sigma N$ coupling interaction
in the process (ii), 
let us make
additional calculations with artificial modification
as follows:
We replace  the $\Lambda N$-$\Sigma N$ coupling part of our
original interaction by the purely central interaction 
and by the purely tensor one without 
changing the other interaction terms,
where the calculated $\Lambda N$ 
phase shifts are kept almost unchanged.
It is found that the effect of the process(ii) 
is quite different between the two cases. Namely,
in the case of the purely central-force coupling,
the contribution is $-1.54$ MeV (attractive) in the
$0^+$ state and   $+0.43$ MeV (repulsive) in the
$1^+$ state.  On the other hand, in the case 
of the purely tensor-force coupling,
it becomes much reduced values, $-0.35 $ MeV for the $0^+$ state 
and $-0.06$ MeV
for the $1^+$ states.
Thus, the resulting energies of $0^+$ and $1^+$ states turn out to be
controlled substantially by the ratio of
the central and the tensor components in the
$\Lambda N$-$\Sigma N$ coupling potential.
Characteristic feature of
our result of the process (ii)
in the $0^+$ state is 
similar to that given by Akaishi {\it et al} \cite{Aka00}.
In the $1^+$ state, however, 
the effect of the process (ii) is much 
different from the result of Ref.\cite{Aka00}:
In our treatment the process (ii) works repulsively or
attractively depending on the ratio of 
the central and the tensor components in the
$\Lambda N$-$\Sigma N$ coupling potential,
while the contribution of
the process (ii)  
 is negligible in Ref. \cite{Aka00} independently of it.
This difference may come from the fact that in the $1^+$ state
of $^4_{\Lambda}$He, 
the probability of the $(NNN)_{\frac{3}{2}} \Sigma$ 
is as large as that of the $(NNN)_{\frac{1}{2}} \Sigma$ 
in our result, but
the $(NNN)_{\frac{3}{2}} \Sigma$ channel 
cannot be  
explicitly included
in  the  coupled two-body model of
$^3{\rm He}+ \Lambda/ \Sigma$ of Ref.\cite{Aka00}.

\section{Summary}
\label{sec:four}

We have developed the calculational method of
four-body bound-state problems so that it becomes possible 
to make precise four-body calculations of 
$^4_{\Lambda}$H and $^4_{\Lambda}$He taking both the
$NNN \Lambda$ and $NNN \Sigma$ channels explicitly
into account using  realistic $NN$ and $YN$ interactions. 
As a result, we succeeded in making clear the
role of $\Lambda$-$\Sigma$ conversion and
deriving the amount of the $\Sigma$-mixing
in $A=4$ hypernuclei quantitatively.
However, our $YN$ interaction employed here is not
sufficient to reproduce the binding energy of the
excited state of $1^+$,
although those of ground states of $^3_{\Lambda}$H, $^4_{\Lambda}$He
and $^4_{\Lambda}$H are in good agreement with
the experimental values.
It is a future problem to explore the
feature of $\Lambda$-$\Sigma$ conversion in $\Lambda$ hypernuclei
with the use of more refined $YN$ interactions through 
systematic study of structure of the heavier hypernuclear systems.

%
\section*{Acknowledgments}

The authors thank Professor Y. Akaishi,  Professor B. F. Gibson 
and Professor T. A. Rijken for helpful
discussions
and Mr. Nogga for discussion on numerical results.
This work was supported by the Grant-in-Aid for Scientific Research in
Priority Areas.


\vfill{\eject}

Table I. Parameters of the central (C), spin-orbit (LS) and tensor (T)
terms of the NSC97f-simulated $YN$ interaction.
Radial form of each term is given by two-range Gaussians.
Range ($\beta$) is in units of fm and strength (V) in units of MeV.  

$$\vbox{
\offinterlineskip
\halign{
%
%
        \enspace\hfil#\hfil\enspace &
        \enspace\hfil#\hfil\enspace &
        \enspace\hfil#\hfil\enspace &
        \enspace\hfil#\hfil\enspace &
        \enspace\hfil#\hfil\enspace &
        \enspace\hfil#\hfil\enspace \cr
%
%
\noalign{\hrule height 0.6pt}
\noalign{\vskip 0.05 true cm} \cr
\noalign{\hrule height 0.6pt}
\noalign{\vskip 0.20 true cm} \cr
&$\beta$ &$V_C(^1E)$  &$V_C(^3E)$  &$V_{LS}$  &$V_T$ \cr
\noalign{\vskip 0.20 true cm} \cr
\noalign{\hrule height 0.6pt}
\noalign{\vskip 0.20 true cm} \cr
$N \Lambda $-$ N \Lambda$ &0.5   &732.08  &1068.8   &1023.8
&$-243.31$  \cr
\noalign{\vskip 0.20 true cm} \cr
&1.2 &$-99.494$  &$-45.490$    &$-17.195$    &$-10.413$  \cr
\noalign{\vskip 0.20 true cm} \cr
\noalign{\hrule height 0.6pt}
\noalign{\vskip 0.20 true cm} \cr
$N \Lambda $-$ N \Sigma$ &0.5 &61.223 &$-770.21$
 &$-19.930$    &287.54  \cr
\noalign{\vskip 0.20 true cm} \cr
 &1.2  &$-15.977$  &68.274  &22.299  &62.438  \cr
\noalign{\vskip 0.20 true cm} \cr
\noalign{\hrule height 0.6pt}
\noalign{\vskip 0.20 true cm} \cr
$N \Sigma$-$ N \Sigma$   & 0.5 &1708.0  &863.76  &544.56   &21.778
\cr
\noalign{\vskip 0.20 true cm} \cr
$(t=1/2)$ &1.2 &80.763  &28.284   &$-19.944$  &$-53.542$  \cr
\noalign{\vskip 0.20 true cm} \cr
\noalign{\hrule height 0.6pt}
\noalign{\vskip 0.20 true cm} \cr
$N \Sigma $-$ N \Sigma$ &0.5 &695.39  &$-181.08$
 &$-462.31$   &333.05  \cr
\noalign{\vskip 0.20 true cm} \cr
$(t=3/2)$ &1.2  &$-109.37$ &23.282  &0.0023  &22.234  \cr
\noalign{\vskip 0.15 true cm} \cr
\noalign{\hrule height 0.6pt}
\noalign{\vskip 0.05 true cm} \cr
\noalign{\hrule height 0.6pt}
}}$$

\vfill{\eject}

Table II. Calculated energies of the $0^+$ and $1^+$ states
          of $^4_{\Lambda}$He and $^4_{\Lambda}$H.
          The energies $E$ are measured from the $NNN\Lambda$
          four-body breakup threshold. 
          $\bar{r}_{A - B}$ denotes
          the r.m.s. distances between  particles $A$ and $B$, while 
          $\bar{r}_{A}$ stands for 
          the r.m.s. radius of particle $A$ measured from the c.m. of 
          $3N$.  As for the $^3$He ($^3$H) nucleus, 
          the calculated binding energy
          is $-7.12 \:(-7.77)$ MeV and 
          $\bar{r}_{N-N}=3.10 \:(3.03)$ fm and 
          $\bar{r}_N=1.79 \:(1.75)$ fm.
          
$$\vbox{
\offinterlineskip
\halign{
%
%
        \enspace\hfil#\hfil\enspace &
        \enspace\hfil#\hfil\enspace &
        \enspace\hfil#\hfil\enspace &
        \enspace\hfil#\hfil\enspace &
        \enspace\hfil#\hfil\enspace &
        \enspace\hfil#\hfil\enspace &
        \enspace\hfil#\hfil\enspace &
        \enspace\hfil#\hfil\enspace \cr
%
%
\noalign{\hrule height 0.6pt}
\noalign{\vskip 0.05 true cm} \cr
\noalign{\hrule height 0.6pt}
\noalign{\vskip 0.15 true cm} \cr
&\multispan2  $^4_{\Lambda}$He &\qquad  &\multispan2
$^4_\Lambda$H \cr
\noalign{\vskip 0.15 true cm} \cr
 &\multispan2    {\hrulefill}    &\qquad    &\multispan2
{\hrulefill}   \cr
\noalign{\vskip 0.15 true cm} \cr
$J$  &$0^+$  &$1^+$    &\qquad  &$0^+$  &$1^+$   \cr
\noalign{\vskip 0.15 true cm} \cr
\noalign{\hrule}
\noalign{\vskip 0.15 true cm} \cr
$E$(MeV)  &$-9.40$  &$-7.66$   &\qquad  &$-10.10$  &$-8.36$ \cr
\noalign{\vskip 0.15 true cm} \cr
$E^{\rm exp}$(MeV)  &$-10.11$  &$-8.87$
   &\qquad  &$-10.52$   &$-9.53$  \cr
\noalign{\vskip 0.15 true cm} \cr
$B_\Lambda$(MeV)  &\enskip2.28   &\enskip0.54   &\qquad  &\enskip2.33
  &\enskip0.59     \cr
\noalign{\vskip 0.15 true cm} \cr
$B_\Lambda^{\rm exp}$(MeV)  &\enskip$2.39$     &\enskip$1.15$
 &\qquad  &\enskip$2.04$  &\enskip$1.05$   \cr
\noalign{\vskip 0.15 true cm} \cr
\noalign{\hrule}
\noalign{\vskip 0.15 true cm} \cr
$\bar{r}_{N-N}$ (fm)     &\enskip2.86   &\enskip3.03  &\qquad
&\enskip2.83   &\enskip2.99  \cr
\noalign{\vskip 0.15 true cm} \cr
$\bar{r}_{\Lambda-N}$ (fm)    &\enskip3.77   &\enskip5.74  &\qquad
&\enskip3.75   &\enskip5.70  \cr
\noalign{\vskip 0.15 true cm} \cr
$\bar{r}_{\Sigma-N}$ (fm)     &\enskip2.24   &\enskip2.48  &\qquad
&\enskip2.23   &\enskip2.46  \cr
\noalign{\vskip 0.15 true cm} \cr
$\bar{r}_{N}$ (fm)     &\enskip1.65   &\enskip1.75  &\qquad
&\enskip1.64   &\enskip1.73   \cr
\noalign{\vskip 0.15 true cm} \cr
$\bar{r}_{\Lambda}$ (fm)     &\enskip3.39   &\enskip5.47  &\qquad
&\enskip3.37   &\enskip5.43   \cr
\noalign{\vskip 0.15 true cm} \cr
$\bar{r}_{\Sigma}$ (fm)     &\enskip1.67   &\enskip1.81  &\qquad
&\enskip1.66   &\enskip1.80   \cr
\noalign{\vskip 0.15 true cm} \cr
\noalign{\hrule height 0.6pt}
\noalign{\vskip 0.05 true cm} \cr
\noalign{\hrule height 0.6pt}
}}$$

\vfill{\eject}

Table III.  The $S$-, $P$- and $D$-state and 
total probabilities of the $(NNN)_{\frac{1}{2}} \Lambda$,
$(NNN)_{\frac{1}{2}} \Sigma$ and
$(NNN)_{\frac{3}{2}} \Sigma$
channels in
the $0^+$ and $1^+$ states of (a)$^4_{\Lambda}$He and
$^4_{\Lambda}$H.
$(NNN)_t$ denotes three nucleons whose isospins are coupled to
$t$.

\vskip 0.30 true cm

\hspace*{-0.5 cm}(a) $^4_{\Lambda}$He
$$\vbox{
\offinterlineskip
\halign{
%
%
        \enspace\hfil#\hfil\enspace &
        \enspace\hfil#\hfil\enspace &
        \enspace\hfil#\hfil\enspace &
        \enspace\hfil#\hfil\enspace &
        \enspace\hfil#\hfil\enspace \cr
%
%
\noalign{\hrule height 0.6pt}
\noalign{\vskip 0.05 true cm} \cr
\noalign{\hrule height 0.6pt}
\noalign{\vskip 0.15 true cm} \cr
 &$S$ (\%)  &$P$ (\%)
&$D$ (\%)  &Total (\%) \cr
\noalign{\vskip 0.05 true cm} \cr
\noalign{\hrule height 0.6pt}
\noalign{\vskip 0.10 true cm} \cr
$0^+$       &    &
  &  & \cr
\noalign{\vskip 0.10 true cm} \cr
$(NNN)_{\frac{1}{2}} \Lambda$   &89.32     &0.08   
&8.52     &97.92  \cr
\noalign{\vskip 0.10 true cm} \cr
$(NNN)_{\frac{1}{2}} \Sigma$   &0.84     &0.04   
&1.16    &2.04  \cr
\noalign{\vskip 0.10 true cm} \cr
$(NNN)_{\frac{3}{2}} \Sigma$   &0.01   &0.01  
&0.02   &0.04 \cr
\noalign{\vskip 0.10 true cm} \cr
\noalign{\hrule height 0.6pt}
\noalign{\vskip 0.10 true cm} \cr
$1^+$   & & & &    
   \cr
\noalign{\vskip 0.10 true cm} \cr
$(NNN)_{\frac{1}{2}}  \Lambda$   &90.38    &0.07   
&8.52    &98.97  \cr
\noalign{\vskip 0.10 true cm} \cr
$(NNN)_{\frac{1}{2}} \Sigma$    &0.10     &0.01   
&0.40  &0.51   \cr
\noalign{\vskip 0.10 true cm} \cr
$(NNN)_{\frac{3}{2}} \Sigma$   &0.09    &0.00  
&0.43   & 0.52  \cr
\noalign{\vskip 0.10 true cm} \cr
\noalign{\hrule height 0.6pt}
\noalign{\vskip 0.05 true cm} \cr
\noalign{\hrule height 0.6pt}
}}$$

\hspace*{-0.5 cm}(b) $^4_{\Lambda}$H
$$\vbox{
\offinterlineskip
\halign{
%
%
        \enspace\hfil#\hfil\enspace &
        \enspace\hfil#\hfil\enspace &
        \enspace\hfil#\hfil\enspace &
        \enspace\hfil#\hfil\enspace &
        \enspace\hfil#\hfil\enspace \cr
%
%
\noalign{\hrule height 0.6pt}
\noalign{\vskip 0.05 true cm} \cr
\noalign{\hrule height 0.6pt}
\noalign{\vskip 0.15 true cm} \cr
 &$S$ (\%)  &$P$ (\%)
&$D$ (\%)  &Total (\%) \cr
\noalign{\vskip 0.05 true cm} \cr
\noalign{\hrule height 0.6pt}
\noalign{\vskip 0.10 true cm} \cr
$0^+$       &    &
  &  & \cr
\noalign{\vskip 0.10 true cm} \cr
$(NNN)_{\frac{1}{2}} \Lambda$   &89.27     &0.08   
&8.54     &97.88  \cr
\noalign{\vskip 0.10 true cm} \cr
$(NNN)_{\frac{1}{2}} \Sigma$   &0.86     &0.04   
&1.19    &2.08  \cr
\noalign{\vskip 0.10 true cm} \cr
$(NNN)_{\frac{3}{2}} \Sigma$   &0.01   &0.01  
&0.02   &0.04 \cr
\noalign{\vskip 0.10 true cm} \cr
\noalign{\hrule height 0.6pt}
\noalign{\vskip 0.10 true cm} \cr
$1^+$   & & & &    
   \cr
\noalign{\vskip 0.10 true cm} \cr
$(NNN)_{\frac{1}{2}}  \Lambda$   &90.34    &0.07   
&8.54    &98.95  \cr
\noalign{\vskip 0.10 true cm} \cr
$(NNN)_{\frac{1}{2}} \Sigma$    &0.10     &0.01   
&0.41  &0.52   \cr
\noalign{\vskip 0.10 true cm} \cr
$(NNN)_{\frac{3}{2}} \Sigma$   &0.09    &0.00  
&0.44   & 0.53  \cr
\noalign{\vskip 0.10 true cm} \cr
\noalign{\hrule height 0.6pt}
\noalign{\vskip 0.05 true cm} \cr
\noalign{\hrule height 0.6pt}
}}$$

\vfill{\eject}

Figure caption 

Fig. 1 Jacobian coordinates for all the rearrangement channels of
$NNN\Lambda (\Sigma)$ system. Three nucleons are to be
antisymmetrized.

Fig. 2 Calculated energy levels of (a)$^4_{\Lambda}$He
and (b)$^4_{\Lambda}$H.
The channels successively included are 
(i) $(NNN)_{\frac{1}{2}}\Lambda$, 
(ii) $(NNN)_{\frac{1}{2}}\Sigma$  and
(iii) $(NNN)_{\frac{3}{2}}\Sigma$ where
the isospin of the three nucleons is coupled
to $t=$1/2 or 3/2.  Energy is measured from the
$^3$He$+\Lambda$($^3$H$+\Lambda$) threshold.

Fig. 3 Calculated energy levels of $^4_{\Lambda}$He
with the AV8, Bonn A and Minnesota potentilas.

Fig. 4  Correlation functions (two-body densities) 
of the $NN$, $\Lambda N$ 
and $\Sigma N$ pairs in the $0^+$ state of 
$^4_{\Lambda}$He together with that for 
the NN pair in $^3$He. Here, the correlation function
of $\Sigma N$ pair has been multiplied by factor 2 so as to
see the behavior of this function clearly.

Fig.5 Calculated one-body ddensities of 
$N$, $\Lambda$ and $\Sigma$ particles in the
$0^+$ state off $^4_{\Lambda}$He.
Volume integrals of the densities are
1.0, 0.98 and 0.02 for $N$, $\Lambda$ and $\Sigma$ particles,
respectively.

Fig. 6 Calculated energy levels of
$^4_{\Lambda}$He for case (i) and case (i)+(ii); here,
(i) denotes the two-body process  and
(ii) denotes the three-body process.

----------------

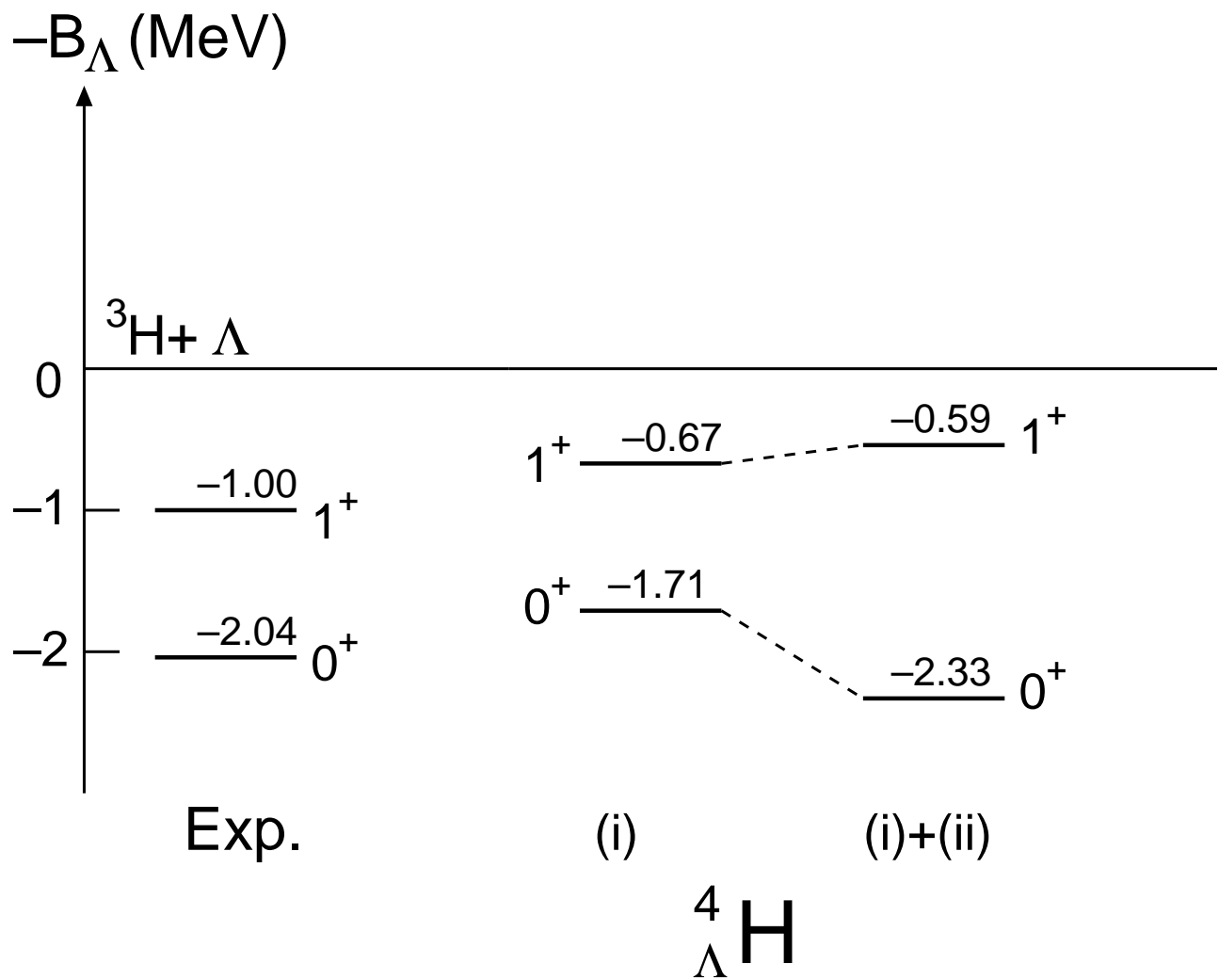
\end{document}